# A Performance-Consistent and Computation-Efficient CNN System for High-Quality Automated Brain Tumor Segmentation

Juncheng Tong and Chunyan Wang

*Abstract*—The research on developing CNN-based fully-automated Brain-Tumor-Segmentation systems has been progressed rapidly. For the systems to be applicable in practice, a good processing quality and reliability are the must. Moreover, for wide applications of such systems, a minimization of computation complexity is desirable, which can also result in a minimization of randomness in computation and, consequently, a better performance consistency. To this end, the CNN in the proposed system has a unique structure with 2 distinguished characters. Firstly, the three paths of its feature extraction block are designed to extract, from the multi-modality input, comprehensive feature information of mono-modality, paired-modality and cross-modality data, respectively. Also, it has a particular three-branch classification block to identify the pixels of 4 classes. Each branch is trained separately so that the parameters are updated specifically with the corresponding ground truth data of a target tumor areas. The convolution layers of the system are custom-designed with specific purposes, resulting in a very simple config of 61,843 parameters in total. The proposed system is tested extensively with BraTS2018 and BraTS2019 datasets. The mean Dice scores, obtained from the ten experiments on BraTS2018 validation samples, are $0.787\pm0.003$, $0.886\pm0.002$, $0.801\pm0.007$, for enhancing tumor, whole tumor and tumor core, respectively, and $0.751\pm0.007$, $0.885\pm0.002$, $0.776\pm0.004$ on BraTS2019. The test results demonstrate that the proposed system is able to perform high-quality segmentation in a consistent manner. Furthermore, its extremely low computation complexity will facilitate its implementation/application in various environments.

## I. INTRODUCTION

BRAIN tumors cause serious brain diseases and brain tumor detection is important for the diagnosis. In general, brain tumor segmentation is to detect and to localize tumors in 3D brain images. As each tumor area can be further segmented into 3 intra-tumoral structures, namely edema (ED), non-enhancing /necrotic tumor (NET) and enhancing tumor (ET), brain tumor segmentation is, ultimately, to classify the voxels of a brain image into 4 classes, i.e., ET, tumor core (TC) comprising ET and NET, whole tumor (WT), and the background.

In the case of brain scanning by Magnetic Resonance Imaging (MRI), four 3D brain images of 4 modalities, i.e., Flair, T2, T1ce and T1, are acquired for each patient case. Each 3D image is usually sliced into a large number of 2D slices. An example of brain image slices of the 4 modalities and the segmentation result approved by neuroradiologists [1] is illustrated in Fig. 1. The brain tumor segmentation is, in fact, a task of multi-class classification of multi-modality data. It requires (i) examination of a large amount of data, and (ii) good knowledge of medical specialists. Hence, manual segmentation is a time-consuming and difficult task, which may lead to lengthy wait for diagnosis results in many patient cases. Developing fully automated brain tumor segmentation system by computer vision is imperative to facilitate timely diagnosis.

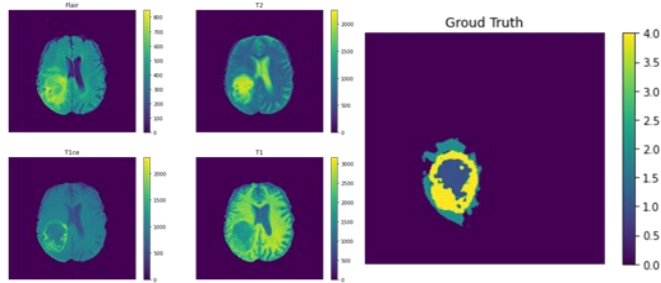

Fig. 1. Example of the four-modality MRI image and its corresponding ground truth of tumor area.

Brain tumor segmentation by computer vision comprises two processes, extraction of the image feature data representing tumor patterns and regrouping the feature data for classification. The feature extraction is done mainly by filtering operations performed sequentially and/or concurrently. Various filters, such as Sobel filters [2] and Gabor filters [3], have been used to extract feature data of different patterns. However, the tumor patterns have a lot of variations, and one cannot apply a large number of filters of handcrafted coefficients to deal with them, which causes a loss of critical information and thus affects the

The work presented in this manuscript is related to the research project of Juncheng Tong's Master's thesis defended on April 13th, 2022 the Department of Electrical and Computer Engineering, Concordia University. More details can be found in the thesis that will soon be available at Spectrum, Concordia University's open access research repository at https://spectrum.library.concordia.ca/cgi/search/advanced.

This work was supported in part by Compute Canada and in part by the Natural Sciences and Engineering Research Council (NSERC) of Canada.
Juncheng Tong is with Gina Cody School of Engineering and Computer Science, Concordia University, Montreal, H3G 1M8, Canada (email: ju_tong@encs.concordia.ca).
Chunyan Wang (Corresponding author) is with Gina Cody School of Engineering and Computer Science, Concordia University, Montreal, H3G 1M8, Canada (email: chunyan@ece.concordia.ca).

quality of the classification in the succeeding process. The feature data regrouping for classification can also be done by means of filtering operations, but it has also the difficulty in manually selecting a large number of suitable filters to correctly classify the variable tumor areas.

Convolutional neural networks (CNN) have been used effectively for various image processing purposes [4]-[8]. As the filter coefficients in a CNN can be determined by means of progressive update in a training process, without the need to choose them manually, one can apply a sufficient number of filters to handle various kinds of feature data. Moreover, as the network is trained with the ground truth data verified by medical specialists, the filters of the network can be set-up and fine-tuned with their knowledge in the topic area.

It should, however, be mentioned that, operating a large number of filters in a CNN requires a huge number of calculations, let alone training the filters, which may be a problem of application in a computation resource restricted environment. Also, while inadequacy of filtering capacity affects the processing quality, excessively use of filters can result in problems, such as performance inconsistency due to the randomness in the training process.

In this paper, we propose a highly reliable CNN system specifically for brain tumor segmentation of MRI brain images. The CNN in the system is custom-designed with the objective that (i) its structure is suitable to process multi-modality input data and to perform multi-class classification, and (ii) its dimensions, i.e., the number of layers and number of filtering kernels per layer, are just-sufficient for the complexity of the task to maximize the computation efficiency and minimize the randomness in the training. To this end, new design approaches are developed to have (a) each filter placed on purpose, (b) the network training optimized, and (c) the computation efficiency and performance consistency maximized. By achieving the objective of the work, it also contributes to the CNN community with the following issues.

- Multi-path feature extraction (FE) block for multi-modality input data. The input data of each modality carry specific feature information and are correlated with the data of the other modalities. The multi-path structure of the block permits each path to be adjusted to optimize the extraction of mono-modality, paired-modality or cross-modality features.
- Decomposition of a complex multi-classification problem into several binary classification problems by a multi-branch classification block. The config of each branch is trained independently to better identify the patterns of a particular tumor area, and the filtering coefficients can be updated separately and specifically using the corresponding ground truth.
- High computation-efficiency and consistency achieved by custom-designing the CNN system. The proposed system, with 61,843 parameters in total, is capable of producing high-quality segmentation results in a consistent manner. The standard deviations of the mean Dice scores of WT, TC and ET, obtained from 10 experiments, are less than 1%.

This paper consists of 5 sections. The related work is presented in Section II. The detailed description of the proposed system is found in Section III. Section IV is dedicated for the presentation of the training details and the experiment results. A conclusion is presented in Section V.

## II. RELATED WORK

U-net structure [9], composed of a contracting path to extract features and an expanding path with skip connection to perform classification, is widely used in developing CNNs for image segmentation. In particular, to segment 3D medical images, the U-net structure has been adapted with different variations.

As the 3D brain images are often seen as sequences of 2D slices, many segmentation systems involve 2D U-net based CNNs, of which the input and ground truth data are 2D samples. In order to use the inter-slice feature information, there are also 3D CNNs operating with 3D input samples. In the implementation of such 3D CNNs, one of the challenging issues is to get training samples of sufficient quality and quantity. In some cases, the limitation of computation resources, including memory usage, can also be an issue. To bypass such issues, varieties of pseudo-3D CNNs have appeared. Some used the combination of 2D and 3D convolutions where 2D ones were for extracting intra-slice features, while 3D ones were for extracting inter-slice features [10]. Others sliced each 3D input image into three orthogonal axes to obtain 3 sequences, then performed 2D convolutions in order to capture spatial information with low dimensional convolutions [11]-[15].

To enhance the computation in convolutional layers, in many U-net based CNNs, multi-convolution blocks were used as the basic computational units. For example, a standard convolution layer was replaced by a residual block [14], [15], inception block [16], [17], or dense block [10], [18]-[20]. In some U-net designs, the skip connections were enhanced by employing attention gates [15], [21], or squeeze-and-excitation blocks [20], [22]. In the case of Unet++ [23], simple skip connections were replaced by a network of convolutional units.

The variations of U-net can also be in network structure. One could use multiple contracting paths in order to process the input data of different modalities. To handle the input data of each modality independently, instead of single one in the original U-net, four contracting paths were used to avoid the false-adaptation between the modalities [24]. There's also a U-net with 2 contracting paths to extract features from flair-t2 pair and t1ce-t1 pair respectively in order to capture the discriminative patterns between brain tumor areas and normal brain areas [25].

Since the brain tumor detection is a fine classification of pixels in a hierarchical structure of tumor sub-regions, multiple U-nets architecture was used where the result of the former U-net was utilized as the prior information for the next U-net to perform finer segmentation process [26]-[29]. For example, three independent U-nets were used to detect WT, TC and ET, successively [27]. One could also cascade multiple U-nets in a single network to reduce the training time and computation cost [28], [29].

However, it should be noted that the high performance of the CNN system is often achieved at the cost of large computation resources, which limits its application in a computation resource restricted environment and increases the risk of performance inconsistency. Therefore, the work presented in this paper aims at custom-designing a CNN system that is able to perform high-quality brain tumor segmentation in a consistent manner with very low computation cost.

### III. PROPOSED SYSTEM

The input of the proposed system consists of four 3D MRI images, corresponding to the 4 modalities, namely Flair, T2, T1ce and T1. Each 3D image is sliced into 2D slices. Thus, the voxels in 3D images become pixels in 2D space. The output is a sequence of slices, in which each pixel is labeled to be in one of the 4 classes, i.e., ED, NET, ET and the background.

The proposed system consists of the following three parts.
1. Pre-processing. It is to exclude the excessive margins and evident tumor-free slices, which reduces the regions of non-interest and the data volume to be processed.
2. CNN. It's the main part of the proposed system to perform brain tumor segmentation. It involves a multi-path feature extraction block to takes fullest advantage of the multi-modality input, a bottleneck to perform tumor localization, and a multi-branch classification block to optimize the segmentation.
3. Refinement block. It aims at correcting falsely labeled enhancing tumor pixels to improve segmentation quality.

#### A. Pre-processing and Refinement Block

The pre-processing and refinement blocks are placed to facilitate the operations in the CNN.

The pre-processing block is to reduce the volume of data to be processed in the CNN, by removing some of the regions that are apparently out of the regions of interest, from the 4 series of slices in each patient case. Two kinds of regions of non-interest are targeted.
- Excessive margins in each slice. In the case of samples from BraTS datasets, one can crop each slice to reduce its dimensions from $240 \times 240$ to $200 \times 168$ pixels, cutting off more than 40% of the input data for the CNN, without losing any pixels in the brain areas.
- Tumor-free slices located in the two ends of each sequence. In the case of Brats samples, the first fifteen slices and the last twelve slices do not contain any tumor areas and are thus removed.

The removal of the 2 kinds of regions of non-interest by the pre-processing block results in a significant data reduction. In terms of BraTS samples, it can reach to 52%, which helps not only to reduce the computation burden in the succeeding CNN, but also to increase the information density of its input data.

The refinement block is to verify if any of the pixels placed, by the CNN, in the ET areas are falsely classified. As a tumor area is a 3D object, the pixels of the area are expected to appear in a certain number of slices and the total number of these pixels should be above a certain threshold. In this block, if an ET area appears in less than 6 slices or the total number of the pixels in the area is less than 1000, the pixels are considered as misclassified and the area will be relabeled as NET.

#### B. CNN

The CNN proposed in this paper is the major part of the system for the brain tumor segmentation. The emphasis in the design of the network is on the performance consistency and computation efficiency, i.e., maximizing the processing quality while minimizing the computation cost and randomness.

The detailed architecture of the proposed CNN is shown in Fig. 2. It consists of the following three blocks.
1. Feature extraction (FE) block. It has three-path convolutions, which are dedicated to extracting different features from the four-modality input.
2. Bottleneck. It is to generate feature maps representing tumor locations of low resolution.
3. Three-branch classification block. Each of the branches is made to identify the pixels of a specific kind of tumor areas.

##### 1) Feature Extraction Block

The function of the FE block is to extract comprehensively image features from 2D MRI image slices of the 4 modalities, namely Flair, T2, T1ce and T1. A set of 4 slices from the 4 modalities is illustrated in the first row of Fig. 3. The brain images are acquired by the 4 modalities to enhance various kinds of pathological tissues to facilitate diagnosis, and the intensity range of different modalities may not be uniformed. Hence, the input data need an appropriate normalization in order to facilitate the following filtering operations.

The aforementioned normalization will be applied to the 3D data of each modality. The mean and standard deviation is calculated from the pixels in the brain areas of all input slices, and those in the background with no region of interest are excluded. As a result, the mean and standard deviation obtained can better represent the distribution of data within the brain area, and consequently, certain features in the brain areas can be enhanced and the contrast of image can be improved as shown in Fig. 3.

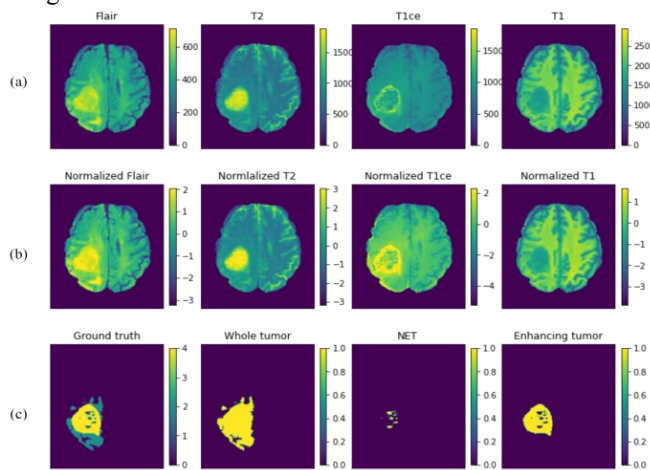

Fig. 3. (a) Input slices of the 4 modalities, (b) normalization results, and (c) ground truth images.

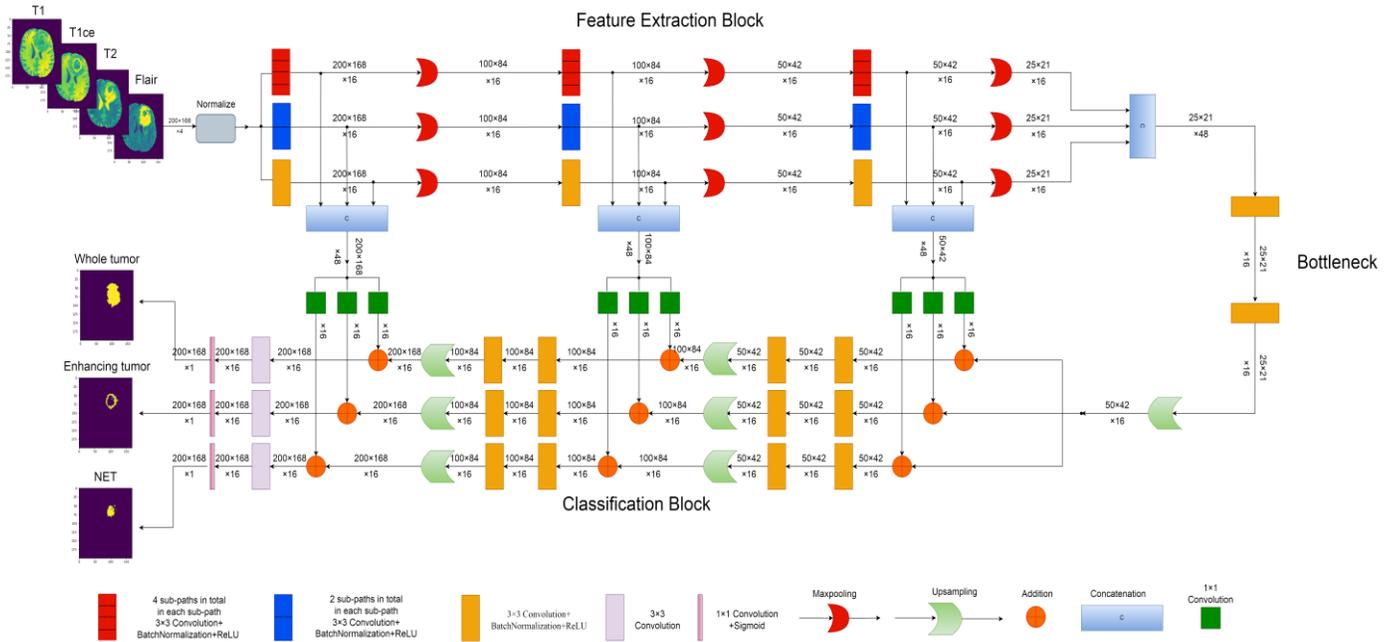

Fig. 2. Detailed diagram of the proposed CNN.

The purpose of the comprehensive feature extraction is to capture the data representing critical pathological patterns from brain images. Considering the different emphasis on the varieties of tissues, the input data of each of the modalities need to be filtered individually in order to make good use of their feature characters. However, as the data of all the 4 modalities are acquired from the same source and therefore correlated, they also need to be processed collectively.

It is known, on one hand, that the patterns of the whole tumor region are distinguishable in Flair and T2, and one can get more whole tumor features by means of convolutions applied to these 2 modalities combined. On the other hand, the pair of T1 and T1ce display more differences between the patterns in tumor core and those in the rest of brain areas, and filtering them together can produce high-quality feature data representing tumor cores and enhancing tumors.

Taking the above-mentioned points into consideration, the FE block is designed with 3 paths, as shown in Fig. 2. The input data of all the 4 modalities are applied to all of the 3 paths, and get convolved 3 times in each path. The input data are organized, however, into 3 different groupings so that the 3 paths can produce the following 3 kinds of features.

- Mono-modality features. The upper path of the FE block has 4 sub-paths, in each of which the input data of mono-modality are processed in the 3 successive layers of convolution-maxpooling operations. Each of the convolutions has 4 kernels, producing adequately 4 data maps per modality. The 4 sub-paths generate, in total, 16 mono-modality feature maps from the input data of Flair, T2, T1 and T1ce.
- Paired-modality features. In the mid path, there are 2 sub-paths, where the input data of Flair-T2 are applied to one sub-path for whole tumor features, and those of T1-T1ce are applied to the other for tumor core and enhancing tumor features. Each convolution has 8 kernels, considering that it involves the data of 2 modalities. There are 16 feature maps generated by the 2 sub-paths.
- Cross-modality features. There is no separated sub-path in the lower path of the FE block. The 3 successive convolution-maxpooling operations are applied to the data of all the 4 modalities combined. This path produces 16 cross-modality feature maps.

A maxpooling operation following each convolution is performed to enlarge the receptive field and to increase the information density, which, nevertheless, reduces the resolution of the feature maps. The 3-path feature extraction block generates, in total, 48 maps of low-resolution (less than 1000 pixels per map), representing, comprehensively, features of cross-modality, individual modality, and Flair-T2 pair & T1-T1c pair.

*2) Bottleneck*

As shown in Fig. 2, the bottleneck is located between the 3-path FE block and the 3-branch classification block. Its function is to transform the input data, i.e., 48 low-resolution feature maps, into data maps indicating the potential tumor locations in order to perform coarse segmentation.

The function of the block is implemented by means of two simple convolution layers. The number of kernels per layer is modestly 16 to minimize the risk of introducing randomness, with a view to achieving a good consistency of the network's performance. Since the input data maps are of low resolution and convolutions do not increase the resolution. The output maps can carry the information of potential tumor locations, without details of tumor areas. The information will be used in the classification block, presented in the following subsection.

*3) Classification Block*

The classification block is to classify all the pixels into 4 classes, i.e., WT, ET, NET, and the background. The prime input of this block is a set of low-resolution data maps produced by the bottleneck.

The basic processing elements in this block are convolution, upsampling and skip connection. The purpose of the upsampling, by means of bilinear interpolation, is to restore, gradually, the initial image dimension. Because the fine classification needs detailed image information which is lost due to the three max-pooling operations in the feature extraction block, the feature maps produced before each maxpooling operation are brought in by means of the skip connections. They are scaled with learnable coefficients, and then used to modulate the upsampled data maps. Two layers of convolutions are employed in the following to fine-tune the modulated signal in order to get precise segmentation result.

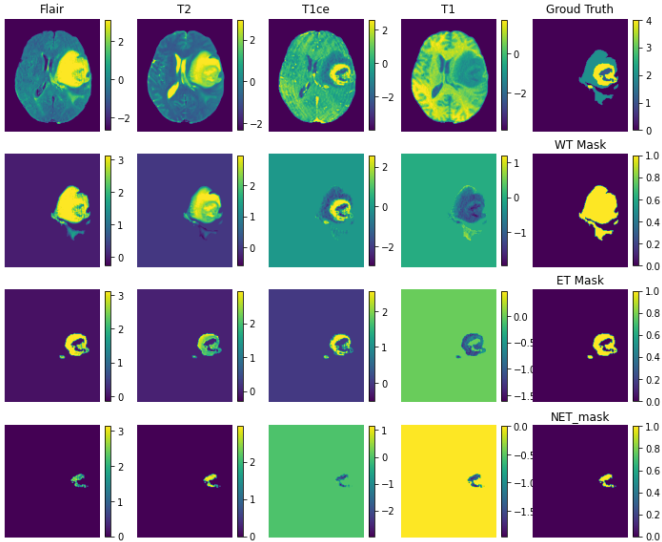

Fig. 4. Different tumor areas have different patterns.

As mentioned earlier, there are 3 kinds of tumor areas needed to be identified, i.e., WT, ET and NET. They are different in sizes, textural patterns, and intensity ranges. The areas of whole tumor, enhancing tumor and NET in a slice of a patient case are illustrated in Fig. 4 as examples. By examining images of patient cases, one can have the following observations.
- Whole tumor areas include various pathological tissues. The have relatively large dimensions. Although the texture patterns have a lot of variations inside the areas, they are different from those outside the areas. Identifying whole areas from a brain image is essentially to distinguish the patterns of pathological tissues from those of the normal brain structures.
- The other kinds of tumor areas, namely ET, NET and edema, have patterns different from one another. If a set of filters is made to detect one kind of tumor areas, it may not be suitable for the others.

In light of above observation, to precisely classify the pixels in the 3 different kinds of tumor areas, different criteria to differentiate image patterns are needed, which may not be easily done by a single filtering path. Hence, the classification block of the proposed system is designed to have 3 branches in parallel, as shown in Fig. 2, dedicated to identifying the pixels in the whole tumor, enhancing tumor, and NET areas, respectively. Each branch is specifically made to differentiate the particular patterns of the certain tumor areas and those in the rest of the brain image. As a result, a complex multi-class classification task is decomposed into several simple binary classification tasks.

The three branches are configured identically, and each has three layers. In each layer, the upsampled data maps are modulated by the feature maps of the same dimensions from the feature extraction block. To optimize the modulation result, the feature maps are appropriately scaled by means of 1x1 convolutions. The modulated data maps are then convolved twice in each layer to perform a precise detection of targeted tumor area.

In order to make each of the 3 branches capable of identifying one specific kind of tumor areas, they are trained independently with the binary masks of the whole tumor, enhancing tumor and NET, respectively. As a result, the filtering coefficients can be adjusted to suit its own pattern detection.

The final output of the classification block is three feature maps indicating the areas of whole tumor, enhancing tumor and NET in the original input.

*4) Network Configuration*

The network configuration is shown is Table I. The total number of parameters is only 61,843, which is merely a small percentage of that of many CNNs in reported literatures.

Table I Details of the Network Configuration.

| Layer | Kernel Size | Input size | Output Size | Kernel Number | Parameters |
|---|---|---|---|---|---|
| 1 | 3×3 | 200×168 | 100×84 | 48 | 1,248 |
| 2 | 3×3 | 100×84 | 50×42 | 48 | 4,272 |
| 3 | 3×3 | 50×42 | 25×21 | 48 | 4,272 |
| 4 | 3×3 | 25×21 | 25×21 | 16 | 9,376 |
| 5 | 3×3 | 25×21 | 50×42 | 48 | 14,304 |
| 6 | 3×3 | 50×42 | 100×84 | 48 | 14,304 |
| 7 | 3×3 | 100×84 | 200×168 | 3 | 7,011 |
| SC1* | 1×1 | 50×42 | 50×42 | 48 | 2,352 |
| SC2 | 1×1 | 100×84 | 100×84 | 48 | 2,352 |
| SC3 | 1×1 | 200×168 | 200×168 | 48 | 2,352 |
| Total | | | | | 61,843 |

* Represents the skip connections.

In general, a performance-consistent, computation-efficient CNN system, combined with pre-processing and refinement block, is custom-designed to perform high-quality brain tumor segmentation based on the analysis of the input data's characters and the output requirements. The proposed network has two highlights.
1. The feature extraction block has 3 paths and each of them is to extract comprehensive features of mono-modality, paired-modality and cross-modality respectively from the multi-modality input data.
2. There are three branches in the classification block, which decomposes a complex multi-class classification problem into three simple binary classification tasks. This structure enables an appropriate training of these branches so that each of them can be adjusted effectively to suit the classification of the pixels of the specific tumor areas.

## IV. Performance Evaluation

The performance evaluation of the proposed system is conducted on the datasets of BraTS 2018 and BraTS 2019. A good number of experiments are carried out to assess the quality of the proposed system in different aspects. The first set of experiments is to evaluate the reproducibility and performance consistency, which is presented in Subsection D. The results of ablation study are found in Subsection E. Subsection F is dedicated for the presentation of the performance comparison between the proposed system and those recently reported in reputed research journals.

### A. Dataset

BraTS 2019 and BraTS 2018 are used to train the system and also for the performance evaluation. In BraTS 2019, there are 335 patient cases in the training pool and additional 125 cases for validation. In BraTS 2018, the number of cases is much smaller and all of them are found in BraTS 2019. The assessment of the validation results is done by means of CBICA Image Processing Portal [30], [31], as the ground truth data are not available in the validation datasets.

A simple data augmentation, by flipping and rotating, is applied to the training samples to increase their number for a better training. Moreover, the training samples are shuffled in each epoch so that they are rearranged differently in batches.

### B. Implementation Details

The hyperparameters determine, to some extent, how the filtering coefficients of the proposed CNN are derived. For the training process, the most important hyperparameters are as follows.
1. Batch size. There are 335 patient cases for training, i.e., 335 * 155 slices per modality. Taking this number of slices and the limit of memory usage into consideration, each batch is made to have 128 slices.
2. Number of epochs. The training requires only 75 epochs for the loss value to reach its minimum level.
3. Learning rate. It is variable with staircase decay. The initial learning rate is 0.005 and will be reduced by half every 15 epochs.
4. Loss function. Dice loss function is applied to calculate the loss in the WT branch. For NET and ET branches, Tversky loss [32] is chosen to reduce the risk of divergence caused by small-size targets.
5. Optimizer. Adam [33] is chosen as the optimizer for the training process.
6. Kernel initialization. The kernel coefficients are initialized with a uniform distribution by he_uniform [34].

Tensorflow is used to implement the proposed system. The total training time is 7 hours for BraTS 2018 dataset and 9 hours for BraTS 2019 dataset with NVIDIA Tesla P100 GPU.

### C. Evaluation Metrics

The segmentation result is mainly evaluated by 4 metrics: Dice score, Sensitivity, Specificity and Hausdorff95. Supposing $P_0$ and $P_1$ are respectively the predicted result of tumor-free and tumor regions, $T_0$ and $T_1$ are the ground truth of tumor-free and tumor regions. Then, the first three metrics can be defined as:

$$Dice(P_1, T_1) = \frac{P_1 \cap T_1}{(P_1 + T_1)/2} \quad (1)$$

$$Sensitivity(P_1, T_1) = \frac{P_1 \cap T_1}{T_1} \quad (2)$$

$$Specificity(P_0, T_0) = \frac{P_0 \cap T_0}{T_0} \quad (3)$$

Hausdorff95 is $95^{th}$ percentile of the maximum distance of a point on the predicted result to the nearest point on the ground truth. Dice score is the most important and comprehensive metric for evaluating brain tumor segmentation result. It measures volumetric overlap between segmentation results and ground truth.

### D. Consistency Study

Due to certain randomness in training a CNN system, one cannot expect such a system to reproduce the exactly same results after each training. Nevertheless, since the proposed system is developed for a medical application, a good reproducibility is required for it to be usable in practice. Hence, it is critical to measure the consistency of its performance to test the reliability of the result.

Ten experiments are conducted, using each of the two datasets, BraTS 2018 and 2019, to measure the consistency, and the results, i.e., the mean and median Dice scores produced in each experiment, are presented in Tables II and III. From the mean Dice scores, presented in the 2nd section in each of the two tables, one can have the following observations concerning the performance consistency.
- The system yields high Dice scores in each experiment. For example, the mean Dice scores of ET in ten experiments are ranged from 0.782 to 0.792 on BraTS 2018 validation dataset, and 0.745 to 0.767 on BraTS 2019 validation dataset. In the case of TC, the mean Dice scores are ranged from 0.792 to 0.813 on BraTS 2018 validation dataset and 0.774 to 0.784 on BraTS 2019 validation dataset.
- The average mean Dice scores of all three tumor areas, obtained from the ten experiments, have a high degree of consistency. The standard deviations are all below 1%.

The median Dice scores of the ten experiments, presented in the 3rd section in each of the two tables, also demonstrate a similar degree of performance excellence and consistency of the proposed system.

The experiment results demonstrate that the proposed system is able to operate in a consistent manner and reproduce fine segmentation results after each retraining, which makes it useful for medical applications. The results also give a confirmation that the measures taken, in the design of the proposed CNN block, to reduce the network randomness are effective. It also proves that a high degree of processing quality and performance consistency can be achieved all together by a custom-designed approach of CNN systems for target-specific applications.

Table II Ten Experiments Results on BRATS 2018 Validation Dataset.

| Exp | Dice-Mean | | | Dice-Median | | |
|---|---|---|---|---|---|---|
| | ET | WT | TC | ET | WT | TC |
| 1 | 0.792 | 0.887 | 0.813 | 0.861 | 0.908 | 0.871 |
| 2 | 0.786 | 0.881 | 0.797 | 0.860 | 0.913 | 0.872 |
| 3 | 0.788 | 0.887 | 0.803 | 0.845 | 0.917 | 0.865 |
| 4 | 0.789 | 0.886 | 0.799 | 0.861 | 0.911 | 0.869 |
| 5 | 0.782 | 0.885 | 0.792 | 0.857 | 0.908 | 0.871 |
| 6 | 0.789 | 0.887 | 0.803 | 0.861 | 0.912 | 0.874 |
| 7 | 0.790 | 0.888 | 0.792 | 0.859 | 0.908 | 0.859 |
| 8 | 0.785 | 0.888 | 0.797 | 0.847 | 0.908 | 0.858 |
| 9 | 0.784 | 0.889 | 0.806 | 0.866 | 0.912 | 0.868 |
| 10 | 0.789 | 0.886 | 0.808 | 0.865 | 0.912 | 0.870 |
| **Best case** | **0.792** | **0.887** | **0.813** | **0.861** | **0.908** | **0.870** |
| **Worst case** | **0.782** | **0.885** | **0.792** | **0.857** | **0.908** | **0.871** |
| **Average** | **0.787** | **0.886** | **0.801** | **0.858** | **0.911** | **0.868** |
| **STDEV** | **0.003** | **0.002** | **0.007** | **0.007** | **0.003** | **0.005** |

Table III Ten Experiments Results on BRATS 2019 Validation Dataset.

| Exp | Dice-Mean | | | Dice-Median | | |
|---|---|---|---|---|---|---|
| | ET | WT | TC | ET | WT | TC |
| 1 | 0.748 | 0.884 | 0.776 | 0.843 | 0.912 | 0.860 |
| 2 | 0.767 | 0.886 | 0.774 | 0.861 | 0.913 | 0.859 |
| 3 | 0.747 | 0.883 | 0.771 | 0.848 | 0.912 | 0.861 |
| 4 | 0.747 | 0.882 | 0.776 | 0.842 | 0.911 | 0.850 |
| 5 | 0.754 | 0.889 | 0.779 | 0.844 | 0.916 | 0.866 |
| 6 | 0.753 | 0.885 | 0.775 | 0.849 | 0.910 | 0.846 |
| 7 | 0.751 | 0.884 | 0.768 | 0.856 | 0.916 | 0.857 |
| 8 | 0.752 | 0.882 | 0.776 | 0.844 | 0.915 | 0.869 |
| 9 | 0.745 | 0.883 | 0.784 | 0.854 | 0.917 | 0.862 |
| 10 | 0.746 | 0.887 | 0.779 | 0.841 | 0.915 | 0.862 |
| **Best case** | **0.767** | **0.886** | **0.774** | **0.861** | **0.913** | **0.859** |
| **Worst case** | **0.745** | **0.883** | **0.784** | **0.854** | **0.917** | **0.862** |
| **Average** | **0.751** | **0.885** | **0.776** | **0.848** | **0.914** | **0.859** |
| **STDEV** | **0.007** | **0.002** | **0.004** | **0.006** | **0.002** | **0.007** |

### E. Ablation Study

The proposed system is designed specifically for brain tumor segmentation and has a number of specific characters, in network configuration and in model parameters, which differentiates it from the other CNN systems for the same task. An ablation study is carried out to assess their effectiveness in processing quality.

The CNN in the proposed system has 4 characters:

1. three-path extraction of mono-modality, paired-modality and cross-modality feature data,
2. classification of 4-class pixels by 3 convolution branches allowing separated training of the filtering coefficients,
3. identification of WT, NET and ET pixels, instead of WT, TC and ET, and
4. minimized numbers of kernels in convolution layers.

Two sets of 4 tests are performed on BraTS 2018 and BraTS 2019 datasets, respectively, to measure the effectiveness of each of the 4 characters in the segmentation quality. The results are presented in Tables IV and V, and each value presented is a mean score obtained in 3 experiments.

In the first test, the 3-path FE block is replaced by a single path involving 3 layers of 48 kernels per layer, and the rest of the network remains unchanged. This replacement, while increasing the total number of parameters from 61.843K to 95.971K, results in a decrease of Dice scores in all the tumor categories. In particular, the ET score is reduced from 0.787 to 0.775 with the data samples of BraTS 2018. With the data samples of BraTS 2019, the decrease is even more significant, i.e., from 0.751 to 0.708. It has been confirmed that the 3-path convolution block effectively optimizes the feature extraction from the 4-modality brain images, and it is thus able to produce high-density features information that leads to a better segmentation.

The second test is carried out to measure the effectiveness of the 3 branches in the classification block. For this measurement, the system is modified in such a way that the 3 branches are combined into single one and the number of kernels in each layer is tripled to be 48, which increases considerably the number of parameters from 61.843K to 115.667K. This modification makes the segmentation less accurate, as shown by the data presented in the second row in Tables IV and V, and the most noticeable is the decrease of 2% in ET Dice score, on both BraTS 2018 and 2019 datasets. This test proves that in the proposed system, the 3-branch structure for the classification of the pixels in 3 different tumor categories is effectively advantageous in terms of processing quality and computation cost. While classification of multi-class is considered as a complex task, this structure permits each branch to be trained separately and specifically to handle a binary classification of a particular tumor category, which simplifies the task and improves the processing quality.

In the third test, the three classification branches are dedicated to identifying pixels of WT, TC and ET instead of WT, NET and ET. In light of the results from Table IV, the dice score of ET in BraTS 2018 is decreased by 3%. One can infer from this observation that the identification of different tumor areas will also influence the network's performance.

The fourth test is to illustrate that the proposed system, despite its very small number of parameters, is able to perform high-quality segmentation of brain images. The test has been conducted in 2 phases. In phase 1, the proposed system is modified in a such way: the number of kernels in each of the 2 layers of the bottleneck is tripled from 16 to 48, and the number of kernels in each of the 3 branches of the classification block decreases from 48 to 16 over the 3 layers instead of constant

Table IV EXPERIMENTS RESULTS OF THE ABLATION STUDIES ON BRATS 2018 VALIDATION DATASET. THE RESULT OF THE PROPOSED SYSTEM IS SHOWN IN BOLD.

| Tests | Ablation type | Dice ET | Dice WT | Dice TC | Sensitivity ET | Sensitivity WT | Sensitivity TC | Number of parameters |
|---|---|---|---|---|---|---|---|---|
| Test1 | Multi-path FE block | 0.775 | 0.882 | 0.791 | 0.822 | 0.898 | 0.804 | 95.971K |
| Test2 | Multi-branch classification | 0.767 | 0.878 | 0.792 | 0.827 | 0.895 | 0.818 | 115.667K |
| Test3 | Identification areas | 0.757 | 0.883 | 0.799 | 0.824 | 0.904 | 0.833 | 61.843K |
| Test4.1 | Number of parameters | 0.782 | 0.887 | 0.794 | 0.843 | 0.903 | 0.817 | 280.419K |
| Test4.2 | Number of parameters | 0.782 | 0.883 | 0.803 | 0.832 | 0.892 | 0.827 | 241.187K |
| **Proposed CNN** | | **0.787** | **0.886** | **0.801** | **0.837** | **0.900** | **0.816** | **61.843K** |

Table V EXPERIMENTS RESULTS OF THE ABLATION STUDIES ON BRATS 2019 VALIDATION DATASET. THE RESULT OF THE PROPOSED SYSTEM IS SHOWN IN BOLD.

| Tests | Ablation type | Dice ET | Dice WT | Dice TC | Sensitivity ET | Sensitivity WT | Sensitivity TC | Number of parameters |
|---|---|---|---|---|---|---|---|---|
| Test1 | Multi-path FE block | 0.708 | 0.881 | 0.771 | 0.731 | 0.877 | 0.779 | 95.971K |
| Test2 | Multi-branch classification | 0.732 | 0.882 | 0.773 | 0.762 | 0.889 | 0.784 | 115.667K |
| Test3 | Identification areas | 0.742 | 0.885 | 0.777 | 0.787 | 0.888 | 0.772 | 61.843K |
| Test4.1 | Number of parameters | 0.752 | 0.886 | 0.768 | 0.791 | 0.888 | 0.761 | 280.419K |
| Test4.2 | Number of parameters | 0.750 | 0.883 | 0.777 | 0.792 | 0.892 | 0.773 | 241.187K |
| **Proposed CNN** | | **0.751** | **0.885** | **0.776** | **0.796** | **0.891** | **0.783** | **61.843K** |

This modification tends to improve the processing quality while the number of parameters is increased from 61.843 k to 280.419k. The results of this test phase are presented in the 4[th] row in Tables IV and V. One can observe that, despite that the number parameters are quadrupled, the segmentation quality, measured by Dice and sensitivity scores, is not improved. In Phase 2, the number of kernels of each convolution is doubled from 16 to 32, making the total number of parameters increase to 241.187K. The test results, illustrated in the 5[th] row, do not show any benefit generated from more filtering kernels. All the data produced in these 2 phases demonstrate that the configuration and the usage of parameters in the proposed system are the most appropriate to optimize the segmentation quality under the current circumstances.

### F. Performance Comparison

The performance of the proposed system is compared, in terms of segmentation quality and computation complexity, with that of those recently reported in reputed journals. The comparison results are presented in Tables VI and VII.

In the two tables, the results of the proposed system are presented in the bottom rows. The Dice and Sensitivity scores are generated by the ten experiments specified in Subsection D. One can have the following two observations.

In terms of segmentation quality, the performance of the proposed system is among the best. Its Dice and Sensitivity scores are higher than that of many other CNN systems reported in research journals, with validation samples of both BraTS 2018 and 2019 datasets.

In terms of computation cost, the proposed system is the best, standing ahead of the others by a considerable distance. The entire convolutional network has only 61,843 parameters, while most existing systems need multi-millions. If the number of voxels of each patient case is 240×240×155×4, the number of Flops to complete the test is merely 146G.

The test results demonstrates that the proposed system is able to perform brain tumor segmentation of excellent quality in a consistent manner, and this performance is achieved at a computation cost that is only a small fraction of the average amount of the others. It confirms the effectiveness of the specific characters in the design. They altogether make the proposed system operate with a high computation efficiency, low randomness, and reliable performance.

### V. CONCLUSION

In this paper, a performance-consistent and computation-efficient, CNN system is proposed for brain tumor segmentation. It's custom-designed on the basis of the analysis of input signal and output requirements. It consists of three parts, (i) a multi-path feature extraction block to extract comprehensive features from the multi-modality input, (ii) a bottleneck to perform tumor localization and coarse segmentation, and (iii) a classification block with three branches to detect different tumor areas separately, where each branch will be trained independently so that the parameters in each branch can be updated accordingly to suit the detection of specific tumor patterns in order to to improve the segmentation accuracy.

Table VI COMPARISON OF THE PROPOSED SYSTEM WITH OTHER TOP-PERFORMANCE CNN SYSTEMS ON BRATS 2018 VALIDATION DATASET. * INDICATES THAT THE NUMBER OF PARAMETERS IS NOT GIVEN IN THE ORIGINAL PAPER AND IT'S AN ESTIMATION BASED ON THE NETWORK CONFIG FROM THE PAPER.

| | Dice | | | Sensitivity | | | Number of parameters |
|---|---|---|---|---|---|---|---|
| | ET | WT | TC | ET | WT | TC | |
| Huang et al. 2021 [35] | 0.717 | 0.801 | 0.759 | 0.829 | 0.962 | 0.800 | N.A. |
| Ben naceur et al. 2020 [18] | 0.732 | 0.860 | 0.733 | 0.740 | 0.838 | 0.702 | 181k |
| Liu et al. 2021 [36] | 0.767 | 0.898 | 0.834 | N.A. | N.A. | N.A. | 3.34M |
| Sun et al. 2021 [37] * | 0.771 | 0.900 | 0.795 | 0.769 | 0.904 | 0.751 | 27.42M |
| Zhang et al. 2020 [21] * | 0.772 | 0.872 | 0.808 | N.A. | N.A. | N.A | 7.5M |
| Rehman et al. 2021 [38] * | 0.773 | 0.894 | 0.826 | N.A. | N.A. | N.A | 31.4M |
| Zhang et al. 2020[39] | 0.782 | 0.896 | 0.824 | N.A. | N.A. | N.A | 363k |
| Best score of the proposed system | **0.792** | **0.887** | **0.813** | **0.851** | **0.901** | **0.820** | **61.834k** |
| Average score of the proposed system | **0.787** | **0.886** | **0.801** | **0.837** | **0.900** | **0.816** | **61.834k** |
| Stddev of the proposed system | **0.003** | **0.002** | **0.007** | **0.007** | **0.003** | **0.005** | **61.834k** |

Table VII COMPARISON OF THE PROPOSED SYSTEM WITH OTHER TOP-PERFORMANCE CNN SYSTEMS ON BRATS 2019 VALIDATION DATASET. * INDICATES THAT THE NUMBER OF PARAMETERS IS NOT GIVEN IN THE ORIGINAL PAPER AND IT'S AN ESTIMATION BASED ON THE NETWORK CONFIG FROM THE PAPER.

| | Dice | | | Sensitivity | | | Number of parameters |
|---|---|---|---|---|---|---|---|
| | ET | WT | TC | ET | WT | TC | |
| Di Ieva, et al. 2021 [40] * | 0.675 | 0.870 | 0.711 | 0.702 | 0.823 | 0.670 | 28.03M |
| Rehman et al. 2021 [38] * | 0.708 | 0.869 | 0.775 | N.A. | N.A. | N.A | 31.4M |
| Zhang et al. 2020 [21] * | 0.709 | 0.870 | 0.777 | N.A. | N.A. | N.A | 7.5M |
| Huang et al. 2021 [35] | 0.730 | 0.827 | 0.788 | 0.798 | 0.967 | 0.801 | N.A. |
| Ali et al. 2021 [41] | 0.740 | 0.880 | 0.810 | 0.770 | 0.890 | 0.790 | N.A. |
| Liu et al. 2021 [36] | 0.759 | 0.885 | 0.851 | N.A. | N.A. | N.A | 3.34M |
| Sun et al. 2020 [37] * | 0.761 | 0.890 | 0.779 | 0.767 | 0.883 | 0.762 | 27.42M |
| Best Score of the proposed system | **0.767** | **0.886** | **0.774** | **0.800** | **0.889** | **0.784** | **61.843K** |
| Average score of the proposed system | **0.751** | **0.885** | **0.776** | **0.796** | **0.891** | **0.783** | **61.834K** |
| Stddev of the proposed system | **0.007** | **0.002** | **0.004** | **0.006** | **0.002** | **0.007** | **61.834K** |

The performance of the proposed system is evaluated on both BraTS 2018 and BraTS 2019 datasets. The results of a good number of experiments demonstrate that the proposed system is able to perform a high-quality segmentation in a consistent manner. The comparison with those recently reported in reputed journals shows that the segmentation quality of the proposed system is among the best while the computation cost is the lowest, merely a small percent of those of other systems.